\documentclass[aps,prl,twocolumn,showpacs,preprintnumbers,amsmath,amssymb]{revtex4}

\newcommand\Tr{{\rm Tr}}
\newcommand\be{\begin{eqnarray}}
\newcommand\ee{\end{eqnarray}}

\begin{document}


\title{Renormalizable Non-Metric Quantum Gravity?}

\author{Kirill Krasnov}
  \affiliation{School of Mathematical Sciences, University of Nottingham, Nottingham, NG7 2RD, UK \\
and \\ Perimeter Institute for Theoretical Physics, Waterloo, N2L 2Y5, Canada}

\date{v2: May 15, 2006}

\begin{abstract} We argue that four-dimensional quantum gravity may be essentially renormalizable 
if one relaxes the assumption of metricity of the theory. We work with Plebanski formulation 
of general relativity in which the metric (tetrad), the connection, and the curvature are 
all independent variables and the usual relations among these quantities are valid only on-shell. One 
of the Euler-Lagrange equations of this theory ensures its metricity. We show that quantum 
corrected action contains a counterterm that destroys this metricity property, and that no other 
counterterms appear, at least, at the one-loop level. The new term in the action is akin to a 
curvature-dependent cosmological ``constant''.
\end{abstract}


\maketitle

Quantum gravity is known to be non-renormalizable. However, its (non-)renormalizability properties have 
been explored mostly in the original Einstein-Hilbert second-order formulation. 
One may argue that the first-order formulation of gravity that uses tetrad instead of the metric 
is more fundamental as one has to employ it if gravity is to be coupled to fermions. In this formulation
there arises a new independent field - the connection $A^{IJ}$, and
the tetrad $\theta^I$ is used instead of the metric. Both $A^{IJ}$ and $\theta^I$ are one-forms,
and the indices $I, J$ run over $0,1,2,3$. The action of the first order gravity is
\be\label{action-first-order}
S[\theta,A]=\frac{1}{8\pi G}\int_M \epsilon^{IJKL} (\theta^I \theta^J F^{KL} + 
\frac{\Lambda}{4}\theta^I \theta^J \theta^K \theta^L).
\ee
Here $F=dA + (1/2)[A,A]$ is the curvature of the Lorentz group Lie algebra valued spin connection $A$,
$\Lambda$ is (a multiple of) the cosmological constant, and the wedge product of all the forms is assumed.
The renormalizability of quantum gravity in 2+1 dimensions is easiest to establish precisely in
such first-order formulation \cite{Witten:1988hc}. One could therefore expect to get some new
insight into the renormalization properties of 3+1 gravity by using the above action as a starting point.
However, one immediate encounters a problem that there is no term in (\ref{action-first-order})
quadratic in the fields that could be interpreted as kinetic.  
Such a term arises only if one assumes the background to be constant, which makes the usual tricks
of the background field method unavailable and the whole perturbative expansion obscure. It is
doubtful that such a background dependent perturbation theory could lead to any new insight.

Another formulation of general relativity was proposed by Plebanski \cite{Plebanski}. The main idea
is that, in addition to the tetrad-like and connection fields, there is a new field that on-shell
becomes identified with the Weyl part of the curvature tensor. Importantly, in this formulation gravity
becomes a non-metric theory - instead of the tetrad one-forms one uses certain new two-forms that
become related to the metric only on-shell. In this paper we will use the original Plebanski self-dual
formulation. For Plebanski-type formulation without the self-dual split see \cite{Capovilla:1991qb}. The action of
the theory is given by
\be\label{action-Plebanski}
S[B,A,\Psi]=\frac{1}{2\pi iG} \int_M B^a F^a + \frac{1}{2}(\Lambda \delta^{ab}+ \Psi^{ab}) B^a B^b.
\ee
Here $a,b$ are the ${\mathfrak su}(2)$ Lie algebra indices, $\Psi^{ab}$ is a
field that on-shell becomes the Weyl part of the curvature (it is required to 
be symmetric traceless), $B^a$ is a Lie algebra valued 2-form field that 
on-shell becomes expressed through a tetrad, and $F^a(A)=dA^a + (1/2)f^{abc} A^b A^c$
is the curvature of the connection $A^a$, and $f^{abc}$ are the ${\mathfrak su}(2)$ Lie 
algebra structure constants. In this self-dual formulation all fields are complex-valued, 
which explains the unusual factor of $1/i$ in front of the action. 
Certain reality conditions need to be imposed to get back to real-valued gravity
of particular metric signature, see e.g. \cite{Capovilla:1991qb} for more details on this. 
Varying the action with respect to $\Psi^{ab}$ one gets
\be\label{metricity}
B^a  B^b = \frac{1}{3} \delta^{ab} \delta^{cd} B^c B^d,
\ee
which can be shown to imply that $B^a$ is the self-dual part of the two form
$B^{IJ}:=(1/2) \epsilon^{IJKL} \theta^{[K} \theta^{L]}$ for some tetrad 
$\theta^I$ (there are some discrete
ambiguities that we are ignoring here, see e.g. \cite{Capovilla:1991qb}). The field $\Psi^{ab}$
thus plays the role of a Lagrange multiplier that imposes the metricity 
condition (\ref{metricity}). In turn, varying the action (\ref{action-Plebanski})
with respect to $B^a$ one gets $F^a= - (\Lambda \delta^{ab}+\Psi^{ab})B^b$, which identifies $\Psi^{ab}$
as the self-dual part of the Weyl curvature tensor.

In formulation (\ref{action-Plebanski}) there is a kinetic term $B^a dA^a + (\Lambda/2)B^a B^a$
quadratic in the fields, and thus a perturbative expansion of the usual type is possible 
(see, however, more remarks on this below).
An immediate question is then how to interpret the term containing $\Psi^{ab}$. 
In this paper we will treat the field $\Psi^{ab}$ as an external background field, and
analyze renormalizability properties of gravity in formulation (\ref{action-Plebanski})
in the background of a chosen $\Psi^{ab}$. Our analysis can thus be compared to a treatment of the 
Dirac electron in the background of an external electromagnetic potential. As is well-known, quantum 
fluctuations of the fermionic field induce a kinetic term of the usual form
$F_{\mu\nu} F^{\mu\nu}$ for the electromagnetic potential. 
One of the aims of our analysis is to see if there is any similar
effect occurring for the curvature field $\Psi$ in the theory (\ref{action-Plebanski}). 

Before we describe the results of a detailed perturbative treatment, it is worth discussing the
renormalizability of (\ref{action-Plebanski}) by power counting. The mass dimensions of the
fields are as follows. The field $B^a$ is related
to the tetrad, and is thus dimensionless, while $A^a, \Psi^{ab}$ are the connection and (on-shell)
curvature fields correspondingly and thus have $[A]=1, [\Psi]=2$. The dimension of the cosmological 
constant is $[\Lambda]=2$. It is clear, however, that this choice of the
dimensions is far from natural. The above action cries for
a rescaling of the fields in which the Newton constant $G$ is absorbed into the $B, \Psi$ fields and $\Lambda$.
The new dimensions are $[B]=2, [\Psi]=0, [\Lambda]=0$. Re-written in this form, 
there is no dimensionfull coupling constant in the theory. 
Moreover, there is not even a dimensionless ``gravitational''
coupling constant left in the theory; in a sense, it is now
the curvature field $\Psi$ that plays the role of such coupling. 

The power counting analysis is now straightforward. Ultraviolet (UV) divergences 
will generate all terms of mass dimension
four compatible with the gauge symmetry. Because the mass dimension of the curvature field $\Psi$ is zero, 
it is clear that all powers of this field will get generated. Thus, in addition to the term
$\Psi^{ab} B^a B^b$ already present in (\ref{action-Plebanski}) the terms of the form
\be\label{divergences}
\frac{1}{2}(\Psi^{k_1})^{ab} (\Tr(\Psi^2))^{k_2} \ldots  (\Tr(\Psi^n))^{k_n} B^a B^b 
\ee
will have to be added to the action to absorb the arising UV divergences, each such
term with its own undetermined coupling constant. The counterterms arising are thus of the type
$(curvature)^n$ -- the theory does seem to be as non-renormalizable as in the 
usual quantum gravity.
In the usual perturbative metric-based treatment the divergences proliferate
because the coupling constant of the theory - the Newton constant - has negative mass dimension.
In the the theory under consideration the multitude of divergences is due
to the fact that there is a field of mass dimension zero. 

In spite of the comforting similarities discussed above, the behavior of the two 
theories under renormalization is rather different, as we shall see. For the moment, let us
assume that the counterterms (\ref{divergences}) are the only ones arising - a very
non-trivial assumption which we shall argue must hold for the theory under consideration 
(after some field redefinitions), see more on this below. 
It is then clear that the whole effect of the counterterms is to replace the curvature
field $\Psi^{ab}$ in (\ref{action-Plebanski}) by a non-trivial functional 
$\tilde{\Psi}^{ab}(\Psi)$ depending on many new parameters (coupling constants) . Importantly,
the functional $\tilde{\Psi}^{ab}$ is no longer traceless. Indeed, let us denote its
trace by $\phi(\Psi):={\rm Tr} \tilde{\Psi}(\Psi)$. We can then write:
\be
\tilde{\Psi}(\Psi)^{ab} = \Phi^{ab}(\Psi) + \frac{1}{3} \delta^{ab} \phi(\Psi),
\ee
where $\Phi^{ab}(\Psi)$ is the traceless
part of $\tilde{\Psi}$. The function $\Phi^{ab}(\Psi)$ is a complicated one, known
only as an expansion in powers of $\Psi$ (with undetermined coefficients). However, it
maps a symmetric trace-free tensor $\Psi^{ab}$ into a symmetric trace-free tensor $\Phi^{ab}$, which
appears in the renormalized action. It is therefore clear that the curvature field 
$\Psi^{ab}$ is a bare, non-observable one, and that $\Phi^{ab}$ just
replaces the original field $\Psi^{ab}$ after the renormalization. We can now, at least in principle,
invert the functional $\Phi^{ab}(\Psi)$ to get $\Psi^{ab}=\Psi^{ab}(\Phi)$, and substitute this into 
the trace function to produce a function $\phi(\Phi)$. 
Thus, provided the assumption we made about the divergences of the
theory is correct, the effect of counterterms consists in replacing the bare curvature field
$\Psi$ by the renormalized one $\Phi$, and in the appearance of a new ``trace'' term in the action.
The resulting renormalized action is:
\be\label{action-renorm}
\frac{1}{i} \int_M B^a F^a  + \frac{1}{2}\left(\Lambda \delta^{ab}+ \Phi^{ab} + 
\frac{\delta^{ab}}{3} \phi(\Phi)\right) B^a B^b.
\ee

Before we return to the assumption made above, let us discuss the implications of 
the new term in the (renormalized) action. It is
easy to see that the metricity equations are modified to
\be\label{non-metricity}
B^a B^b = \frac{1}{3} \left( \delta^{ab} - \frac{d \phi(\Phi)}{d \Phi_{ab}} \right) \delta^{cd} B^c B^d.
\ee
This equation no longer implies that the two-form field $B^a$ is of metric (tetrad) origin. 
We see that non-metricity is unavoidable whenever there is non-zero ``curvature'' $\Phi$. 
We have put the word ``curvature'' in the quotation
marks because it is no longer true that the 
field $\Phi^{ab}$ is the Weyl part of the Riemann curvature for the
metric determined by $B^a$ - there is no such metric anymore. Nevertheless, we will continue to refer
to $\Phi$ as the curvature field. 

If the only arising counterterms are those of the type (\ref{divergences}), the class of theories
(\ref{action-renorm}) is closed under the renormalization group flow. These theories are
then still non-renormalizable in the strict sense of the word, as there is an infinite 
number of undetermined couplings - the coefficients in the expansion 
of $\phi(\Phi)$ into (traces of) powers of $\Phi$ - to be fixed by experiment or some
other principle. However, the renormalization group flow preserving
the form of the action (\ref{action-renorm}) is therefore a flow in the space of one scalar
function of a $3\times 3$ symmetric traceless matrix. As there is only 2 scalar invariants that
can be constructed from such a matrix (namely the traces of its square and cube), the renormalization
group acts in the space of complex-valued functions of two complex variables. Thus, in this
formulation of gravity the renormalization group flow becomes much more tractable. It is in 
this sense that the theory can be referred to as ``renormalizable''. There is then 
the hope that the flow can be understood completely. After this is done one would be not far from
an ultraviolet (UV) completion of the theory, e.g. via the asymptotic safety scenario of Weinberg \cite{Weinberg}.
In the present case, this scenario would translate into a conjecture that there exists
a non-trivial UV fixed point of the renormalization group flow given by 
certain function $\phi_{UV}(\Phi)$. We note also that the renormalizability properties 
of the theory under consideration, i.e. the renormalization group flow acting in the space 
of scalar functions is similar to those of symmetry-reduced gravity model studied by Niedermaier, see 
\cite{Niedermaier:2006ns} for a review. Existence of a non-Gaussian UV fixed point in this
model gives hope that the asymptotic safety scenario may also be realized in the 
theory (\ref{action-renorm}).

To demonstrate that (\ref{action-renorm}) has the properties as described above
it remains to be shown that the counterterms (\ref{divergences}) are indeed
the only ones that arise. Before we discuss this any further, let us see what other counterterms could be
expected. First, we should note that the statement that only (\ref{divergences}) get generated is 
true only with the clause ``up to field redefinitions''. 
Indeed, in addition to the terms we already discussed,
the terms of the form $\tilde{\Psi}_{FF}^{ab}(\Psi) F^a F^b$ and $\tilde{\Psi}_{BF}^{ab}(\Psi) B^a F^b$,
where $\tilde{\Psi}_{FF}(\Psi), \tilde{\Psi}_{BF}(\Psi)$ are series in $\Psi$ (which can start with the constant 
term proportional to $\delta^{ab}$), are all of mass dimension four and are 
gauge invariant, and therefore will get generated. 
However, these terms are easily eliminated by a redefinition of the field $B^a$.
Indeed, the change $B^a \to Q^{ab}B^b + H^{ab}F^b$ with $Q,H$ appropriately
chosen eliminates the term quadratic in the curvature and brings the term $B^a F^b$ into its canonical form.
The other potential terms of concern are the ones containing four and two covariant
derivatives of $\Psi$, e.g.
\be\label{divergences-bad}
\Psi^{aa_1} (d_A \Psi)^{a_1 a_2} (d_A \Psi)^{a_2 a_3} (d_A \Psi)^{a_3 a_4} (d_A \Psi)^{a_4 a}, \\
f^{abc} \Psi^{ba_1} (d_A \Psi)^{a_1 a_2} (d_A \Psi)^{a_2 c} F^a,
\ee
as well as terms similar to the last one with $F^a$ replaced by $B^a$. 
Divergences of this type, if exist, would make
the above argument somewhat less convincing. We claim that no such terms arise
in the theory under investigation (see more on this below). One can also worry about 
the appearance of counterterms dependent on the background metric and required 
to fix the various gauge symmetries present in (\ref{action-Plebanski}). 
Such terms may be present if the BRST
algebra one uses in the gauge-fixing procedure is metric dependent. 
Again, to show that this is not the case
one has to resort to a detailed analysis. 

Let us now briefly describe the main points of the analysis we have performed. 
To avoid dealing with the
field $B^a$ redefinitions as described above, we decided to introduce a 
background only for the gauge field $A$
(and of course the curvature field $\Psi$) but not $B$. There is no loss of 
generality in this, as both the
original action (\ref{action-Plebanski}) and the renormalized action with all the counterterms are 
quadratic in $B^a$. The 2-form field can therefore be integrated out. Thus, we are led to consider 
Einstein's gravity in the form
\be\label{action-pure-gauge}
S[A,\Psi]=\frac{i}{2} \int_M (\Lambda \delta^{ab} + \Psi^{ab})^{-1} F^a F^b,
\ee
which has also been discussed in e.g. \cite{Capovilla:1991qb}. As before, we would like to treat $\Psi$
as an external non-fluctuating field, and integrate over fluctuations of the gauge field about
a chosen (arbitrary) background. In doing this, we would like to treat the first, $\Psi$-independent
term as the kinetic one. However, there is an obvious problem with this, 
as this kinetic term is topological and independent of the connection $A$. 
To circumvent the problem we have followed a trick originally due to St\"uckelberg \cite{Stuck},
also used recently in the context of BF YM theory in \cite{Cattaneo:1997eh},
which consists in introducing an extra field in the action, 
in our case a 2-form Lie algebra valued field $G$, so that 
the full action acquires the desired (topological) symmetry. Thus, we consider
the action (\ref{action-pure-gauge}) with $F$ replaced by $F-G$, 
which is invariant under $A\to A+\tau$ (for infinitesimal $\tau$) 
when $G$ transforms as $G\to G+d_A \tau$.
We can then gauge-fix this symmetry using standard BRST methods. The gauge we have used is the 
so-called self-dual gauge in which $G_{\mu\nu}^a$ is set to be a self-dual two-form. 
Our treatment is motivated by works on Donaldson theory, especially \cite{Labastida:1988qb}, see also 
the review \cite{Birmingham:1991ty}. After fixing the gauge and integrating $G$ out one arrives at the 
following simple theory:
\be\label{action-gauge-fixed}
S[A,\Psi] = - \frac{1}{2} \int_M \left( \Lambda \delta^{ab} + \Psi^{ab} \right)^{-1} 
(F^+)^{\mu\nu a} (F^+)_{\mu\nu}^a,
\ee
plus a certain set of ghost terms, essentially the same as those present in the Donaldson theory;
details will appear elsewhere. The only important point for purposes of this letter 
is that the gauge-fixed BRST algebra is independent
of the background metric introduced for fixing the gauge. This guarantees independence of the
correlation functions of the theory from the metric, see \cite{Blau:1990nv} for a detailed
argument to this effect. 

Before we describe results of our perturbative analysis of (\ref{action-gauge-fixed}) a word of
caution is in order. The equivalence of the physical content of the theory (\ref{action-gauge-fixed}) and
the theory (\ref{action-pure-gauge}) may be questioned, similarly to who one can question the 
equivalence of Donaldson theory $(F^+)^a (F^+)^a$ that restricts one to the moduli space of YM instantons,
and a rather trivial theory $F^a F^a$. The available derivations of one theory from the other by
a ``gauge-fixing'' procedure are only ``metaphorical''. The same can be said about the relation between 
the two theories (\ref{action-pure-gauge}) and (\ref{action-gauge-fixed}). It is because of this delicate
point that we cannot immediately extend the results on (\ref{action-gauge-fixed}) to the
theory (\ref{action-pure-gauge}). Further work comparing the physics of both theories is necessary. 

To understand (\ref{action-gauge-fixed}) perturbatively we use the background field method. 
The Feynman rules for the $\Psi$-independent sector are the usual ones for YM theory in the background field. 
The interaction vertices of the $\Psi$-sector are not much more complicated; because of the lack of 
space we will only describe here the basic vertex of the type $\Psi aa$, 
where $a$ is the fluctuation of the gauge field.  This vertex is proportional to the function
$N_{\mu\nu}(p,q) = (p\cdot q) g_{\mu\nu} - q_\nu p_\mu + i \epsilon_{\mu\nu\rho\sigma} p_\rho q_\sigma$,
which satisfies an important projector-like property
\be
\frac{1}{p_2^2} N_{\mu_1 \mu_2}(p_1,p_2) N_{\mu_2 \mu_3}(p_2,p_3) = N_{\mu_1 \mu_3}(p_1,p_3),
\ee
making computations at any order in $\Psi$ manageable. 

The results of perturbative (one-loop) computations 
for the theory (\ref{action-gauge-fixed}) can be summarized as follows. 
First, there are no divergences that would require counterterms containing derivatives of $\Psi$. 
Together with the metric-independence of the BRST
algebra this implies that all counterterms are of type (\ref{divergences})
containing higher powers of $\Psi$. This means that the theory 
(\ref{action-gauge-fixed}) is renormalizable by the usual multiplicative
renormalization of the gauge-field coupling constant $g$ and by redefinition of the field $\Psi$. 
Our claim of renormalizability properties of quantum gravity (\ref{action-renorm}) 
follows from this renormalizability of (\ref{action-gauge-fixed}),
see however our cautionary remarks above. Using the theory (\ref{action-gauge-fixed}) 
we have also computed the beta-function for the lowest order coupling in the expansion of $\phi(\Phi)$
in powers of $\Phi$. We have found a negative beta-function, indicating that this coupling
becomes important in the infra-red. 

Let us conclude with a brief discussion of the main results. We have shown that
the non-metric generalization (\ref{action-renorm}) of Einstein's GR ``gauge-fixed'' in 
a particular way (\ref{action-gauge-fixed}) described above is one-loop renormalizable 
in the sense that the renormalization group flow acts in the space of
functions of two complex variables. It remains to extend the results obtained to arbitrary 
number of loops, as well as clarify the physical meaning of the ``gauge-fixing'' we have used.  
It is also very important to find a more direct, symmetry based argument that would show that no counterterms
containing derivatives of $\Psi$ get generated. The renormalizability we established is 
not of the usual multiplicative type - we had to invoke both multiplicative
renormalization as well as field redefinitions. It is also not of the usual type as there still remains
a large ambiguity in the theory - the function $\phi(\Phi)$ is undetermined, and is 
to be fixed by experiment or some other principle. The UV completion of the theory may be
provided by the asymptotic safety scenario, provided a non-trivial UV fixed point $\phi_{UV}(\Phi)$
exists. 

Let us compare the theory we considered with renormalizable, asymptotically-free 
theories of quantum gravity proposed in the literature, e.g. \cite{Fradkin:1981hx}.
These theories are similar to the one described here in that higher powers of the curvature 
play the key role. Unlike all previous proposals, the theory discussed contains no higher 
derivatives and is thus free from ghosts. 

In this paper the effect of quantum gravitational corrections was argued to be in 
appearance in the renormalized action (\ref{action-renorm}) of a term playing 
the role of a curvature dependent cosmological ``constant''. This interpretation of
the quantum corrected gravity action may shed new light on the 
cosmological constant problem.

\end{document}